# Advantageous grain boundaries of iron pnictide superconductors

Takayoshi Katase[1], Yoshihiro Ishimaru[2], Akira Tsukamoto[2], Hidenori Hiramatsu[1], Toshio Kamiya[1], Keiichi Tanabe[2] and Hideo Hosono[1, 3,*]

[1] Materials and Structures Laboratory, Mailbox R3-1, Tokyo Institute of Technology, 4259 Nagatsuta-cho, Midori-ku, Yokohama 226-8503, Japan

[2] Superconductivity Research Laboratory, International Superconductivity Technology Center, 10-13 Shinonome 1-chome, Koto-ku, Tokyo 135-0062, Japan

[3] Frontier Research Center, S2-6F East, Mailbox S2-13, Tokyo Institute of Technology, 4259 Nagatsuta-cho, Midori-ku, Yokohama 226-8503, Japan

(*) E-mail: hosono@msl.titech.ac.jp

Phone: +81-45-924-5359

Fax: +81-45-924-5339

**Abstract**

High critical temperature superconductors have zero power consumption and could be used to produce ideal electric power lines. The principal obstacle in fabricating superconducting wires and tapes is grain boundaries–the misalignment of crystalline orientations at grain boundaries, which is unavoidable for polycrystals, largely deteriorates critical current density. Here, we report that High critical temperature iron pnictide superconductors have advantages over cuprates with respect to these grain boundary issues. The transport properties through well-defined bicrystal grain boundary junctions with various misorientation angles ($\theta_{GB}$) were systematically investigated for cobalt-doped $BaFe_2As_2$ ($BaFe_2As_2$:Co) epitaxial films fabricated on bicrystal substrates. The critical current density through bicrystal grain boundary ($J_c^{BGB}$) remained high (> 1 MA/cm$^2$) and nearly constant up to a critical angle $\theta_c$ of ~9º, which is substantially larger than the $\theta_c$ of ~5º for $YBa_2Cu_3O_{7-\delta}$. Even at $\theta_{GB} > \theta_c$, the decay of $J_c^{BGB}$ was much smaller than that of $YBa_2Cu_3O_{7-\delta}$.

Grain boundary (GB) engineering of high critical temperature ($T_c$) cuprate superconductors has been a critical issue in developing practical applications such as superconducting wires and tapes[1], because their superconducting properties heavily depend on the misorientation angle ($\theta_{GB}$) at GBs; therefore, grains in cuprate superconductors must be highly textured to minimize deterioration of the critical current density ($J_c$) across misoriented GBs. In a representative cuprate superconductor, $YBa_2Cu_3O_{7-\delta}$ (YBCO), a fundamental study of intergrain $J_c$ in bicrystal GBs ($J_c^{BGB}$) has been performed using several types of bicrystal substrates[2]. Significantly misaligned adjacent grains cause $J_c^{BGB}$ to decay exponentially as a function of $\theta_{GB}$ from 3º to 40º [3]. Therefore, to produce YBCO superconducting tapes with a high $J_c$, it is necessary to insert well-aligned buffer layers with a small distribution of in-plane misalignment $\Delta\phi <$ 5º on polycrystalline metal substrates using the ion-beam-assisted deposition (IBAD) technique[4] or rolling-assisted biaxially textured substrates (RABiTS)[5]. Although recent progress in buffer-layer technology has established $\Delta\phi <$ 5º, which provides a self-field $J_c$ of several MA/cm² at 77 K[6], fabricating such a buffer layer is time-consuming and expensive. The development of new high-$T_c$ and high upper critical magnetic field ($B_{c2}$) superconducting materials with a more gradual $J_c^{BGB}(\theta_{GB})$ dependence allows for a simple and low-cost fabrication process and provides wider flexibility in superconductor power lines and their applications.

Iron pnictide superconductors[7,8] have received increasing interest as a new family of high-$T_c$ superconductors because they have attractive properties such as a high $T_c$ up to 56 K [9] and high $B_{c2}(0)$ well over 100 T [10,11]. The iron pnictide family shares several characteristics with the cuprates, such as a layered crystal structure, superconductivity induced by carrier doping and the presence of competing orders. However, there are some noteworthy differences, such as the metallic nature of normal states of the iron pnictides instead of the antiferromagnetic Mott insulators of the cuprates, smaller anisotropy in superconducting properties, which is primarily due to the multi-pocket structures of their Fermi surfaces, and a highly symmetric order parameter based on $s$-wave instead of a $d$-wave pairing one[12]. Therefore, we expect that these characteristics of iron pnictides would make them more favorable than cuprates with respect to the supercurrent conduction mechanism in misoriented GBs.

We have studied epitaxial films of the $AE$Fe$_2$As$_2$ ($AE$ = Sr and Ba) system[13–18] because they have considerably small anisotropy $\gamma = B_{c2}^{B \perp c} / B_{c2}^{B // c}$ = 1.9–1.5 in superconducting properties, high $B_{c2}(0) > 50$ T and weak thermal fluctuations with a Ginzburg number $G_i$ of $1.7 \times 10^{-4}$ among iron pnictide materials[19, 20]. In particular, cobalt-doped BaFe$_2$As$_2$ (BaFe$_2$As$_2$:Co) appears to have great potential for device applications[15–17,21–24] because it is rather easy to grow films by pulsed laser deposition (PLD) and chemically stable in an ambient atmosphere[15]. It was previously reported that in-field transport properties across bicrystal GBs (BGBs) with 4 different $\theta_{GB}$ = 3°–24° formed in BaFe$_2$As$_2$:Co epitaxial films grown on [001]-tilt SrTiO$_3$ (STO) bicrystal substrates suggested that even low-angle BGB with $\theta_{GB}$ = 6° obstructs supercurrent and the weak-linked BGBs exhibit a similar behavior to YBCO BGBs[22]. In this article, we report comprehensive results on transport properties of BaFe$_2$As$_2$:Co BGB junctions grown directly on insulating bicrystal substrates in the full range of $\theta_{GB}$ from 3° to 45°. Among the results, it is particularly noteworthy that the critical angle $\theta_c$ of ~9° for BaFe$_2$As$_2$:Co is much larger than the previously-reported value and the decay slope is much smaller than that of cuprates. The large $\theta_c$ allows a simpler and lower-cost fabrication process of superconducting tapes. This advantageous GB nature is demonstrated by the high $J_c >$ 1 MA/cm$^2$ of a BaFe$_2$As$_2$:Co superconducting tape fabricated on a polycrystalline flexible metal substrate[25].

**Results**

**BaFe$_2$As$_2$:Co bicrystal grain boundary (BGB) junctions.** BaFe$_2$As$_2$:Co epitaxial films with the optimal Co concentration of 8% were fabricated on [001]-tilt bicrystal substrates of MgO with $\theta_{GB}$ = 3°–45° and (La, Sr)(Al, Ta)O$_3$ (LSAT) with $\theta_{GB}$ = 5°–45° by PLD with a Nd:YAG laser ($\lambda$ = 532 nm) at a substrate temperature of 850 °C[16]. To date, different techniques that employ conductive buffer layers of STO, BaTiO$_3$ [23] or Fe [26] have been proposed to enhance high-quality epitaxial growth. However, we have reported that it is possible to grow high-quality BaFe$_2$As$_2$:Co epitaxial films with high self-field $J_c >$ 1 MA/cm$^2$ directly on insulating MgO and LSAT single-crystalline substrates without any buffer layer, which has been achieved by optimizing the growth conditions[16]. The directly grown BaFe$_2$As$_2$:Co films exhibited an onset $T_c$ of 20.7 K for MgO bicrystal substrates and 21.6 K for LSAT bicrystal substrates with sharp transition widths ($\Delta T_c$) of 1.1 K[17].

Figure 1 (a) illustrates the device structure fabricated on [001]-tilt bicrystal substrates. To perform transport measurements of $J_c^{BGB}$, the BaFe$_2$As$_2$:Co epitaxial films were patterned into 300-μm-long, 8-μm-wide micro-bridge structures. To compare the intergrain $J_c$ ($J_c^{BGB}$) and intragrain $J_c$ ($J_c^{Grain}$), two types of micro-bridges were fabricated: one was a 'BGB junction' that contained a BGB each, and the other was a 'Grain bridge' that did not contain a BGB. The electrical contacts were formed with In metal pads and Au wires. The current–voltage (I–V) characteristics of BGB junctions and Grain bridges were measured by the four-probe method.

**High critical angle of strong-link – weak-link transition.** The correlation between the transport $J_c^{BGB}$ and the $\theta_{GB}$ is the most important index in characterizing the GB properties of superconductors. Figure 1 (b) summarizes the self-field $J_c^{BGB}(\theta_{GB})$ measured at 4 K and 12 K, and Figure 1 (c) shows the $J_c^{BGB}/J_c^{Grain}$ ratio at 4 K. For comparison, the generally accepted average $J_c^{BGB}(\theta_{GB})$ properties of YBCO BGB junctions measured at 4 K and 77 K [2] are also plotted. The $J_c^{Grain}$ for all of the BaFe$_2$As$_2$:Co Grain bridges are greater than 1 MA/cm$^2$ at 4 K and 1.0–0.5 MA/cm$^2$ at 12 K. For the BGB junctions with low $\theta_{GB} \leq \sim 9°$, the $J_c^{BGB}/J_c^{Grain}$ ratio remained almost at unity, indicating that the low-angle BGB junctions do not behave like a weak link. However, with the increase in $\theta_{GB}$ from 9° to 45°, $J_c^{BGB}$ decreases to approximately 5%, which indicates that the transition from the strong link to the weak link occurs at $\theta_c$ of approximately 9°. In the YBCO system, the $J_c^{BGB}$ values at 4 K for $\theta_{GB} < 5°$ were slightly less than the $J_c^{Grain}$ values, and the $J_c^{BGB}$ values for $\theta_{GB} > 5°$ showed a clear weak-link behavior[2, 27]. Although the reported data are somewhat scattered[2], the typical values of $\theta_c$ = 3–5° are almost half the magnitude of those obtained for the BaFe$_2$As$_2$:Co BGB junctions.

**Gentle $J_c^{BGB}$ decay in weak-link regime.** It is well known that the BGB junctions of the cuprates exhibit nearly exponential decay in their $J_c^{BGB}(\theta_{GB})$ curves in the weak-link regime with an empirical equation of $J_{c0}\exp(-\theta_{GB}/\theta_0)$, where $\theta_0$ denotes the characteristic angle. The $J_c^{BGB}(\theta_{GB})$ curves are expressed by $3.0\times10^7\exp(-\theta_{GB}/4.3°)$ at 4 K and $7.0\times10^6\exp(-\theta_{GB}/4.2°)$ at 77 K. The present BaFe$_2$As$_2$:Co BGB junctions also show an exponential decay, approximated as $2.8\times10^6\exp(-\theta_{GB}/9.0°)$ at 4 K (the red line in Fig. 1(b)) and $1.5\times10^6\exp(-\theta_{GB}/8.5°)$ at 12 K (the blue line in Fig. 1(b)). It should be noted that the $\theta_0$ values for the BaFe$_2$As$_2$:Co BGB junctions are twice as large as those for the YBCO BGB junctions, which indicates that the $J_c^{BGB}(\theta_{GB})$ of the BaFe$_2$As$_2$:Co BGB junctions shows a more gradual decrease than that of the YBCO BGB junctions. Consequently, the $J_c^{BGB}$ of the BaFe$_2$As$_2$:Co BGB junction exceeds that of the YBCO BGB junctions at $\theta_{GB} \geq 20°$ at 4 K.

**BGB junction characteristics.** Figure 2 (a) shows the *I–V* characteristics measured at 12 K for 8-μm-wide BGB junctions with $\theta_{GB}$ = 4°, 16° and 45°. The BGB junction with $\theta_{GB}$ = 4° only exhibits a sharp resistivity jump at a large $I_c$ of 50 mA due to the normal state transition. Similar *I–V* characteristics were observed for all of the BGB junctions with $\theta_{GB} \leq 9°$. On the other hand, the *I–V* curves of BGB junctions with $\theta_{GB}$ = 16° and 45° show nonlinear characteristics without hysteresis in the low voltage region. In general, the shapes of the *I–V* curves of BGB junctions depend on the fractions of the Josephson current exhibiting resistively shunted junction (RSJ) behavior ($W_{RSJ}$) and a supercurrent showing flux flow (FF) behavior ($W_{FF}$). A phenomenological model to explain their fractions in *I–V* curves was previously proposed as follows.[28] *I–V* curves are expressed by the combination of the RSJ and the FF behaviors, where the RSJ current follows $I_{RSJ} = \left(\left(\dfrac{V}{R_N}\right)^2 + I_c^2\right)^{1/2}$ and the FF current follows $I_{FF} = I_S - A\exp\left(-\dfrac{V}{V_0}\right)$ ($R_N$ is the normal-state resistance of the barrier, and $I_S$, $A$ and $V_0$ are constants). The fitting results based on this model drawn by the blue lines reproduce the experimental *I–V* curves well. The fractions of the RSJ current $W_{RSJ}$ are approximately 0%, 70% and 100% for the BGB junctions with $\theta_{GB}$ = 4°, 24° and 45°, respectively. The 100% RSJ current is further confirmed by good agreement with the commonly used Ambegaokar-Halperin (AH) model[29] drawn by the red line for the BGB junctions with $\theta_{GB} = 45°$. The RSJ behavior in the I–V curves of the BGB junctions with $\theta_{GB} = 45°$ was observed in the whole temperature range below $T_c$. The junction resistance $R_NA$, where A is the cross-sectional area of the junction, provides information on the nature of the barrier in BGB junctions. The $R_NA$ products were estimated by fitting the above-mentioned model to the experimental I–V curves[28]. The $R_NA$ of the BaFe$_2$As$_2$:Co BGB junctions are $5\times10^{-11}$ $\Omega$cm$^2$ for $\theta_{GB} = 16°$ and $5\times10^{-10}$ $\Omega$cm$^2$ for $\theta_{GB} = 45°$, which are one or two orders of magnitude smaller than those of the YBCO BGB junctions ($6\times10^{-9}$–$8\times10^{-8}$ $\Omega$cm$^2$ for $\theta_{GB} = 16°$–$45°$, respectively). These results suggest that the BaFe$_2$As$_2$:Co BGB junctions work as superconductor–normal metal–superconductor (SNS) junctions. On the other hand, the YBCO BGB junctions generally show hysteretic I–V curves at low temperatures well below $T_c$, indicating relatively large junction capacitances and resistances due to the insulating nature of their junction barriers. In the case of the BaFe$_2$As$_2$:Co BGB junctions, the nonhysteretic curves, even at low temperatures, suggest the metallic nature of the junction barriers.

Figures 2 (b) and (c) show the Josephson junction properties of 3-µm-wide BGB junctions with a small $\theta_{GB} = 16°$ and a large $\theta_{GB} = 45°$, respectively. The left figures show the magnetic field B dependencies of $I_c$ ($I_c$–B) under B applied perpendicular to the film surfaces, and the right figures show the I–V curves of the BGB junctions irradiated with microwaves at a frequency of 1.39 GHz for $\theta_{GB} = 16°$ and 2.0 GHz for $\theta_{GB} = 45°$ measured at 16 K. The $I_c$–B pattern of the BGB junction with $\theta_{GB} = 45°$ is distinct from the ideal Fraunhofer pattern, probably due to the inhomogeneous current distribution along the BGB; however, the junctions exhibit an $I_c$ modulation of almost 100%, which corresponds to the fact that the BGB junction with $\theta_{GB}$ = 45° exhibits the 100% RSJ current. The BGB junction with $\theta_{GB}$ = 16° exhibits an $I_c$ modulation of only 35% due to the excess current attributable to the FF current. Furthermore, both devices clearly show Shapiro steps with periodic current steps. The measured step voltage heights of 2.9 μV for $\theta_{GB}$ = 16° and 4.1 μV for $\theta_{GB}$ = 45° are consistent with the Josephson relation $V_{RF} = \dfrac{nhf_{RF}}{2e}$, where $f_{RF}$ is the frequency of the applied microwaves.

**Atomic structures of BGBs.** Here, we examined the microstructure and the local chemical composition deviation of the BGBs to check the effect of an impurity phase on the weak-link junction behaviors. Figures 3 (a)–(c) show plan-view high-resolution transmission electron microscope (HR-TEM) images of the BaFe$_2$As$_2$:Co BGB junctions with $\theta_{GB}$ = 4°, 24° and 45°, respectively. The [100]-axes of BaFe$_2$As$_2$:Co are indicated by the arrows. Symmetrically tilted junctions were formed in almost the entire region of the BGBs for all of the junctions. In the BGB junctions with $\theta_{GB}$ = 4° and 24°, periodic misfit dislocations at intervals of approximately 5.0 nm for $\theta_{GB}$ = 4° and 1.2 nm for $\theta_{GB}$ = 24° are clearly observed along the BGBs. On the other hand, the BGB with $\theta_{GB}$ = 45° has blurred lattice fringes across the entire region. Using a geometric tilted boundary model, the grain boundary dislocation spacing $D$ is given by $D = (|b|/2)/\sin(\theta_{GB}/2)$, where $|b|$ is the norm of the corresponding Burgers vector[3]. With the lattice constant $a$ = 0.396 nm of BaFe$_2$As$_2$[30], $D$ is estimated to be 5.7 nm and 1.0 nm for $\theta_{GB}$ = 4° and 24°, respectively. The estimated $D$ values are very similar to the $D$ values observed above. For $\theta_{GB}$ = 45°, the $D$ value is estimated to be 0.5 nm. This value is almost the same as the lattice parameter; therefore, we cannot observe periodic misfit dislocations in the BGB with $\theta_{GB} = 45°$ in Fig. 3 (c). Energy dispersive spectroscopy (EDS) line spectra across the BGBs and parallel to the BGBs confirmed that the chemical compositions of the BGBs and the film region are homogeneous, and no secondary phase was observed in the BGB regions.

Next, we discuss the relationship between $\theta_c$ and the BGB dislocation spacing $D$ observed by TEM. For the BaFe$_2$As$_2$:Co BGB junctions, the observed critical angles $\theta_c$ are approximately 9°, which correspond to a $D$ value of approximately 2.8 nm. This is comparable to or slightly larger than the coherence length $\xi_{ab}(T)$ of 2.6 nm at 4 K estimated from the reported $\xi_{ab}(0\ K) = 2.4$ nm for BaFe$_2$As$_2$:Co[19]. The above relationship supports the notion that strong current channels still remain between the dislocations when $\theta_{GB}$ is below $\theta_c$, while coherent superconducting current cannot pass through the BGBs at $\theta_{GB} > \theta_c$ and behaves like a weak link.

Note that there would be other factors that affect the GB transport properties. For instance, the dislocation cores formed along BGBs can produce residual strains, which have been considered to play one of the major roles in current blocking at BGBs of cuprates[31,32]. This possibility would help us to obtain a more informative insight into the weak-link behavior in iron pnictide superconductors; however, further microstructure and strain analyses are necessary to evaluate the strain field and discuss their effects.

**In-field characteristics of BGB junctions.** To investigate the weak-link behavior in a $B$, $J_c^{BGB}(B)$ values for the BGB junctions with $\theta_{GB} = 3°$–$45°$ were measured at $B$ up to 9 T applied parallel to the $c$-axis. Figure 4 (a) shows the $J_c^{BGB}(B)$ curves measured at 4 K, and the inset figure shows a magnified view in the low $B$ region up to 0.2 T. The $J_c^{Grain}$ measured for the Grain bridge on the 3° MgO bicrystal substrate is also plotted by the black squares. For the BGB junctions with $\theta_{GB}$ = 3° and 4°, the $J_c^{BGB}(B)$ values are almost indistinguishable from those of the $J_c^{Grain}(B)$ curves, and a reduction in $J_c^{BGB}$ is not observed. The other BGB junctions with larger $\theta_{GB}$ show more rapid decreases than those of the $J_c^{Grain}(B)$, even in a low magnetic field. The $J_c^{BGB}(B)$ of the BGB junctions with large $\theta_{GB}$ = 24° and 45° decrease sharply to 2% and 0.8% of the $J_c^{Grain}(B)$ upon the application of 0.1 T. For the BGB junctions with $\theta_{GB}$ = 8° and 11°, the $J_c^{BGB}(B)$ curves show an intermediate behavior between the strongly linked and weakly linked states, where the rapid decrease in $J_c^{BGB}(B)$ at < 1 T can be attributed to the weak pinning force of the flux trapped at the dislocation cores despite the existence of strong channels with almost the same width as the coherence length $\xi_{ab}(T)$.

**Metallic nature of high-angle BGB junctions.** The temperature dependence of $J_c^{BGB}$ ($J_c^{BGB}(T)$) provides more information about the nature of the barriers formed in BGBs. In general, $J_c$ of a SNS junction follows a quadratic temperature dependence. On the other hand, a superconductor–insulator–superconductor (SIS) junction shows a linear dependence on $T$. Figure 4 (b) shows the $J_c^{BGB}(T)$ from $T_c$ down to 4 K for BGB junctions with high $\theta_{GB}$ = 16°–45°. All of the $J_c^{BGB}(T)$ curves of the BaFe$_2$As$_2$:Co BGB junctions show a quadratic temperature dependence as described by

$I_c = I_0 \left(1 - \dfrac{T}{T_c}\right)^2 \dfrac{\kappa d}{\sinh(\kappa d)}$, which is the de Gennes' theory based on a conventional proximity effect in the dirty limit[33]. This relation is indicated by the orange lines in the figure, where $d$ is the barrier thickness and $\kappa^{-1}$ is the decay length for normal metal. In particular, the BGB junctions with $\theta_{GB}$ = 30° and 45° have this quadratic relationship with temperature, which is more clearly confirmed in the $J_c - \left(1 - \frac{T}{T_c}\right)^2$ plots in the inset figure. For the BGB junctions with $\theta_{GB} = 16°$ and $24°$, $J_c^{BGB}(T)$ deviates downward from the quadratic dependence and assumes a linear relation at temperatures lower than 10 K and 8 K, respectively. This result can be explained by the long junction limit with $w/\lambda_J > 4$ because the critical currents become so large that the Josephson penetration depth $\lambda_J$ becomes smaller than the junction width $w$. The dependencies of the $J_c^{BGB}(T)$ curves of the BaFe$_2$As$_2$:Co BGB junctions are distinctly different from those reported for the YBCO BGB junctions because the latter closely follow the linear relation $\alpha\left(1 - \frac{T}{T_c}\right)$ over a wide temperature range below $T_c$[34].

**Discussion**

The doubly-larger $\theta_c$ and the much gentler slope decay than those of YBCO BGB junctions make it easier to produce high $J_c$ BaFe$_2$As$_2$:Co superconducting tapes because the formation of a buffer layer with $\Delta\phi \leq 9°$ is much easier than those used for the cuprates, which require much smaller $\Delta\phi$ of $< 5°$, but such buffer layers have been achieved only by a few groups using an additional MgO or CeO$_2$ buffer layers[35,36]. Therefore, the large $\theta_c$ allows us to use a simpler and lower-cost production process of superconducting tapes, and the iron pnictide superconducting tapes would find practical applications under a higher magnetic field if further improvement in $J_c$ will be achieved.

The powder-in-tube technique has been rather progressing as an alternative technology for superconducting wires also in iron pnictides[37]; however, their $J_c$ values still remains at ~10$^4$ A/cm$^2$ [38] due probably to existence of large angle GBs with $\theta_{GB}$ much greater than $\theta_c = 9°$. The grain boundary issue in iron pnicides will be largely relaxed by the present finding. Actually, we recently succeeded in obtaining high transport $J_c = 3.5$ MA/cm$^2$ with a BaFe$_2$As$_2$:Co biaxially textured thin film on a polycrystalline Hastelloy tape with an IBAD-MgO textured buffer layer[25].

In conclusion, we fabricated high-quality BaFe$_2$As$_2$:Co films with large $J_c^{Grain}$ on bicrystal substrates with the entire range of $\theta_{GB} = 3–45°$ and comprehensively examined the grain boundary nature of the iron pnictide. The primary point clarified by the present study is that the BaFe$_2$As$_2$:Co BGB junctions exhibit a large $\theta_c$ of ~9°. The low-angle BGBs with $\theta_{GB} \leq 9°$ consist of long-period dislocation cores and, therefore, $J_c^{BGB}$ is similar to $J_c^{Grain}$; while the high-angle BGBs show a weak-link behavior with a gradual decay of $J_c^{BGB}(\theta_{GB})$ expressed by the exponential equation of $2.8\times10^6\exp(-\theta_{GB}/9.0°)$. Such grain boundary natures together with the high $B_{c2}(0)$ make the iron pnictides to be more promising materials for application to high $J_c$ superconducting tapes.

**Methods**

**Fabrication of BaFe$_2$As$_2$:Co epitaxial films on bicrystal substrates.** BaFe$_2$As$_2$:Co epitaxial films were fabricated by pulsed laser deposition (PLD) on [001]-tilt bicrystal substrates of MgO with $\theta_{GB}$ = 3º–45º and also of LSAT with $\theta_{GB}$ = 5º–45º. A Nd:YAG laser (wavelength: 532 nm, INDI-40, Spectra-Physics) typically used for epitaxial growth of iron pnictide films[13,39] with a repetition rate of 10 Hz on the PLD target of a high-purity BaFe$_{1.84}$Co$_{0.16}$As$_2$ polycrystalline disk was used as the excitation source[16]. Films with 250–350 nm in thickness were grown at a temperature of 850 °C and the thickness of each film was measured precisely with a surface profiler. The base pressure in our PLD chamber was ≤ 1×10$^{-6}$ Pa, and film deposition was carried out in a vacuum at approximately 10$^{-5}$ Pa. The BaFe$_2$As$_2$:Co epitaxial films grown under these conditions showed high $J_c$ values of 1–4 MA/cm$^2$ at 4 K, which were confirmed by *I–V* characteristic measurements with a 1-μV/cm criterion[17].

**Transport properties through BGB.** The BaFe$_2$As$_2$:Co films were patterned using photolithography and Ar-ion milling into 300 μm-long and 8 μm-wide micro-bridge structures (Fig. 1(a)) to perform four-terminal *I–V* measurements of the $J_c^{BGB}$ across the BGB and of the $J_c^{Grain}$ not across the BGB and under magnetic fields perpendicular to the film surface. The critical current ($I_c$) and the asymptotic junction resistance ($R_NA$) were estimated from the *I–V* characteristics[28].

**Microstructure and chemical composition analysis of BGB.** The microstructures around the BGBs were examined by plan-view high-resolution transmission electron microscopy (HR-TEM). The TEM samples were prepared by a focused-ion-beam (FIB) micro-sampling technique in which the area near the BGBs was mechanically cutout, and that area was only thinned by FIB technique. All of the operations were performed in a high-vacuum chamber. The chemical composition of the bulk film and the BGBs was analyzed by energy dispersive X-ray spectroscopy (EDS) with a spatial resolution of approximately 1 nm.

**Acknowledgment**

This work was supported by the Japan Society for the Promotion of Science (JSPS), Japan, through the "Funding Program for World-Leading Innovative R&D on Science and Technology (FIRST) Program".

**Author contribution**

All authors equally contributed to the work presented in this paper.

**Additional information**

Competing financial interests: The authors declare no competing financial interests.

T. Katase et al.

**Figure legends**

**Figure 1 | Critical current density ($J_c$) versus misorientation angle ($\theta_{GB}$). (a)** Device structure of the BGB junctions and Grain bridges. The upper rectangular solid is an enlargement at the BGB junction. **(b)** Transport intergrain critical current density $J_c^{BGB}$ in a self-field measured in BaFe$_2$As$_2$:Co BGB junctions grown on [001]-tilt bicrystal substrates of MgO (indicated by open symbols) and LSAT (closed symbols) with $\theta_{GB}$ = 3°–45°. The $J_c^{BGB}$ of the BaFe$_2$As$_2$:Co BGB junctions were taken at 4 K (red symbols) and 12 K (blue symbols), and the red and blue solid lines are fitted to the empirical equation of $J_c^{BGB} = J_{c0}\exp(-\theta_{GB}/\theta_0)$. The average data of the YBCO BGB junctions taken at 4 K and 77 K are also indicated by the green and orange dashed lines, respectively[2], for comparison. **(c)** The ratio of the intragrain $J_c$ ($J_c^{Grain}$) and $J_c^{BGB}$ to $\theta_{GB}$ = 0°–25° at 4 K. Open and closed symbols show the ratios of samples on MgO and LSAT bicrystals, respectively. The dashed green line shows the result of the YBCO BGB junctions[2].

**Figure 2 | Josephson junction properties. (a)** Intergrain current–voltage (*I*–*V*) characteristics for the BaFe$_2$As$_2$:Co BGB junctions with $\theta_{GB}$ = 4°, 16° and 45° grown on MgO bicrystal substrates measured at 12 K. The experimental data are plotted by the black open circles. The red dotted lines indicate the fitted results by the A – H model based on the RSJ like transport[29]. The blue lines show the fit to the *I*–*V* curves with the phenomenological model combining the RSJ behavior and the FF behavior[28]. **(b)** and **(c)** Josephson junction properties at 16 K of **(b)** the BGB junctions with $\theta_{GB}$ = 16° and **(c)** the BGB junctions with $\theta_{GB}$ = 45°. The left figures show the magnetic field (*B*) dependence of critical current (*I*$_c$). The *B* was changed along the direction of the

horizontal arrows. The right figures show the Shapiro steps under microwave irradiations at a frequency of 1.39 GHz for $\theta_{GB} = 16°$ and 2.0 GHz for $\theta_{GB} = 45°$.

**Figure 3 | [001] plan-view HR-TEM images of the BaFe$_2$As$_2$:Co BGB junctions on MgO bicrystal substrates.** The misorientation angle $\theta_{GB}$ = **(a)** 4°, **(b)** 24°, and **(c)** 45°. The directions of BGB junctions are indicated by dashed arrows and the [100]-axes are symmetrically tilted from the BGB lines. The misfit dislocations are marked by the down-pointing arrows. Each horizontal bar indicates 5 nm scale.

**Figure 4 | Magnetic field and temperature dependence of $J_c^{BGB}$ for the BaFe$_2$As$_2$:Co BGB junctions. (a)** $J_c^{BGB}(B)$ curves of BaFe$_2$As$_2$:Co BGB junctions with $\theta_{GB}$ = 3°–45° grown on MgO bicrystal substrates measured at 4 K. The intragrain $J_c$ ($J_c^{Grain}$) values measured in the Grain bridge on a 3° MgO bicrystal substrate are also plotted by closed symbols. The inset shows a magnified plot in the low magnetic field up to 0.2 T. **(b)** $J_c^{BGB}(T)$ for the BGB junctions with high $\theta_{GB}$ = 16°–45° grown on MgO bicrystal substrates measured from $T_c$ down to 4 K. The orange lines show the variation of $J_c$ with temperature as predicted from de Gennes' theory[33]. The inset shows a linearized plot of the quadratic temperature dependence for $\theta_{GB}$ = 30° and 45°.

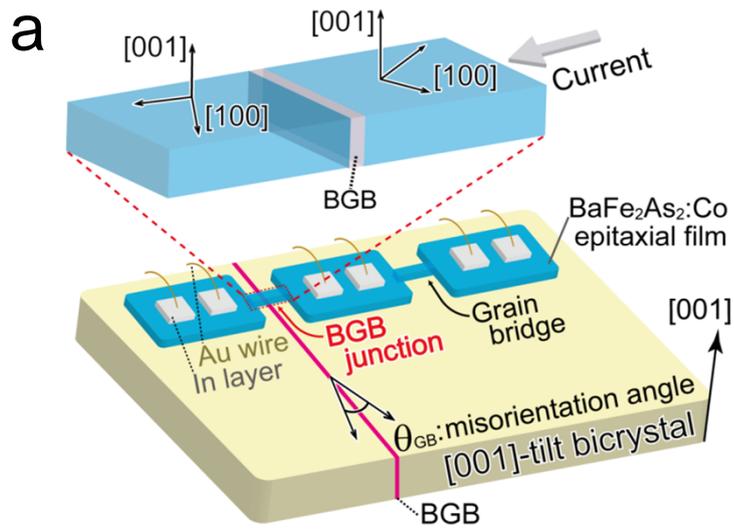
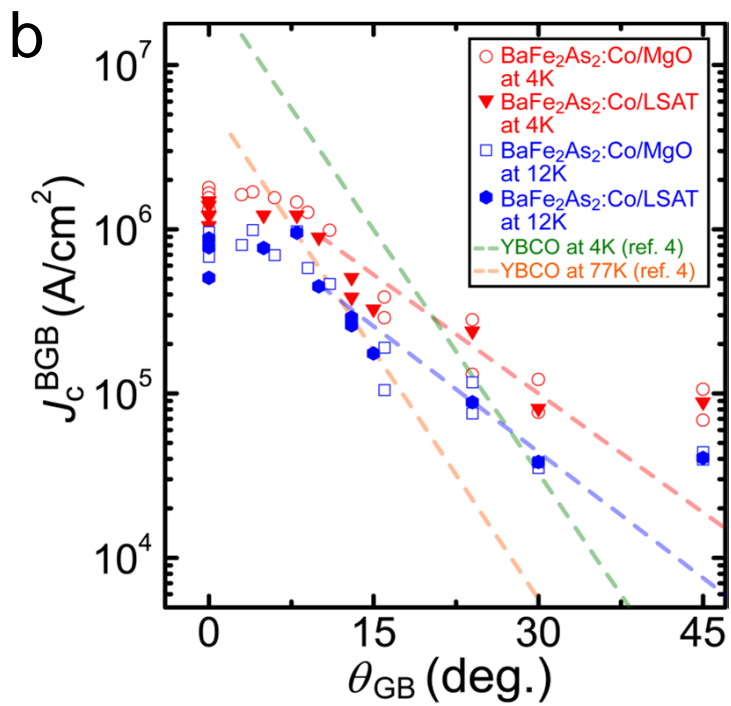
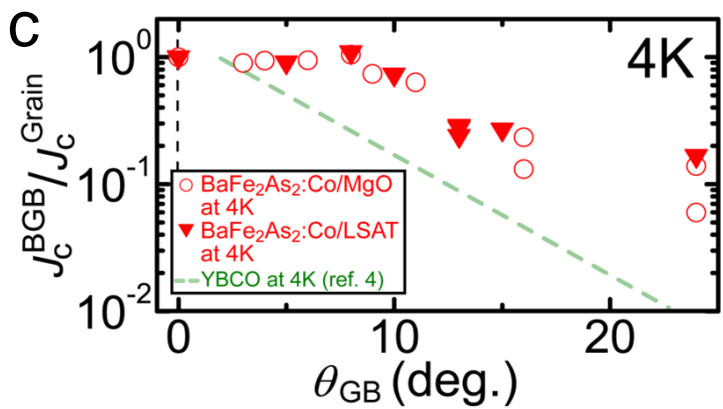

Figure 1

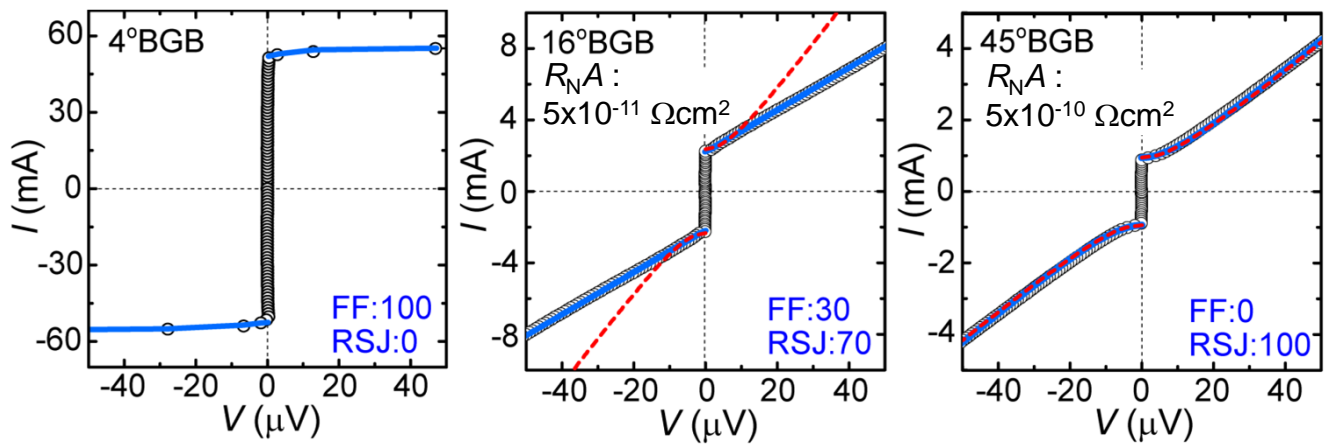
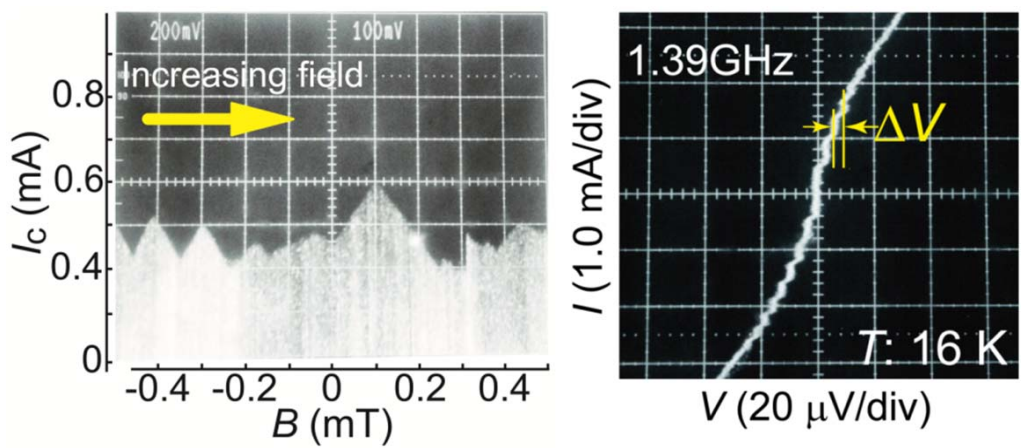
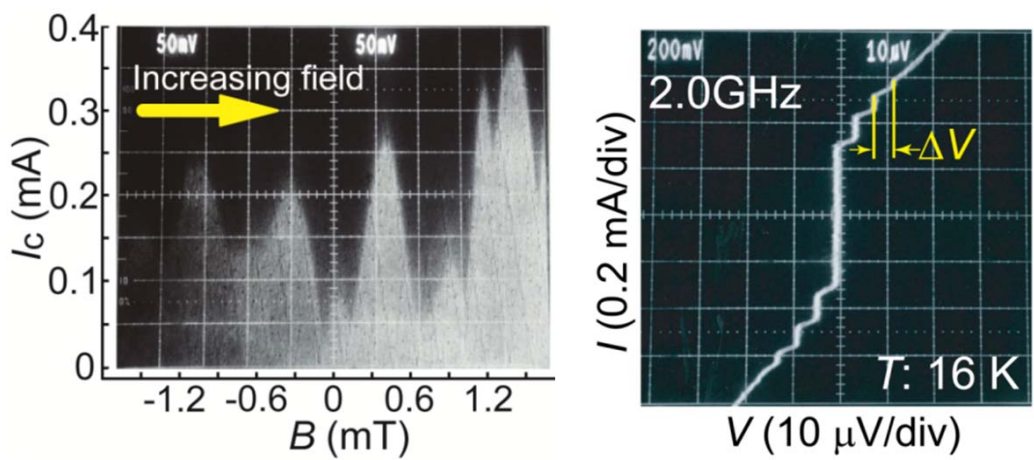

Figure 2

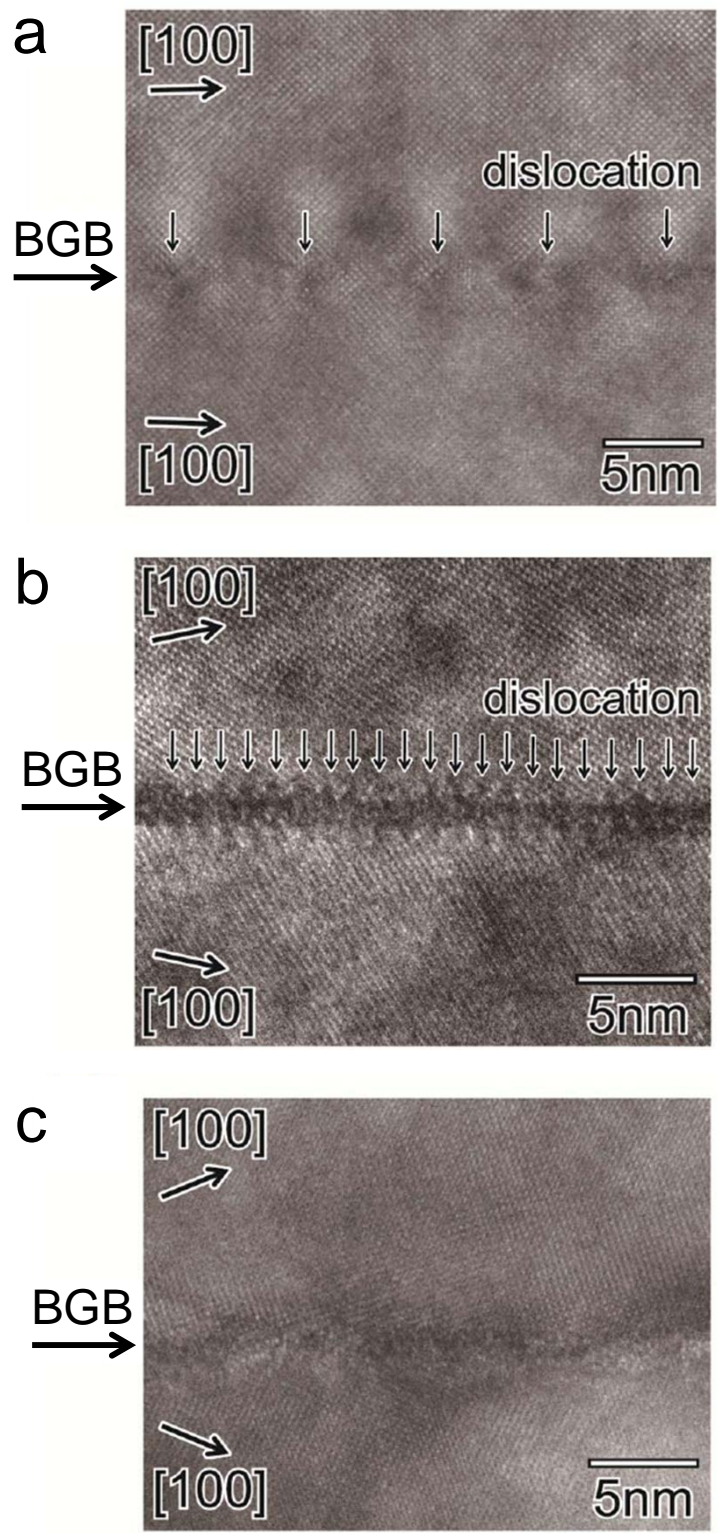

Figure 3

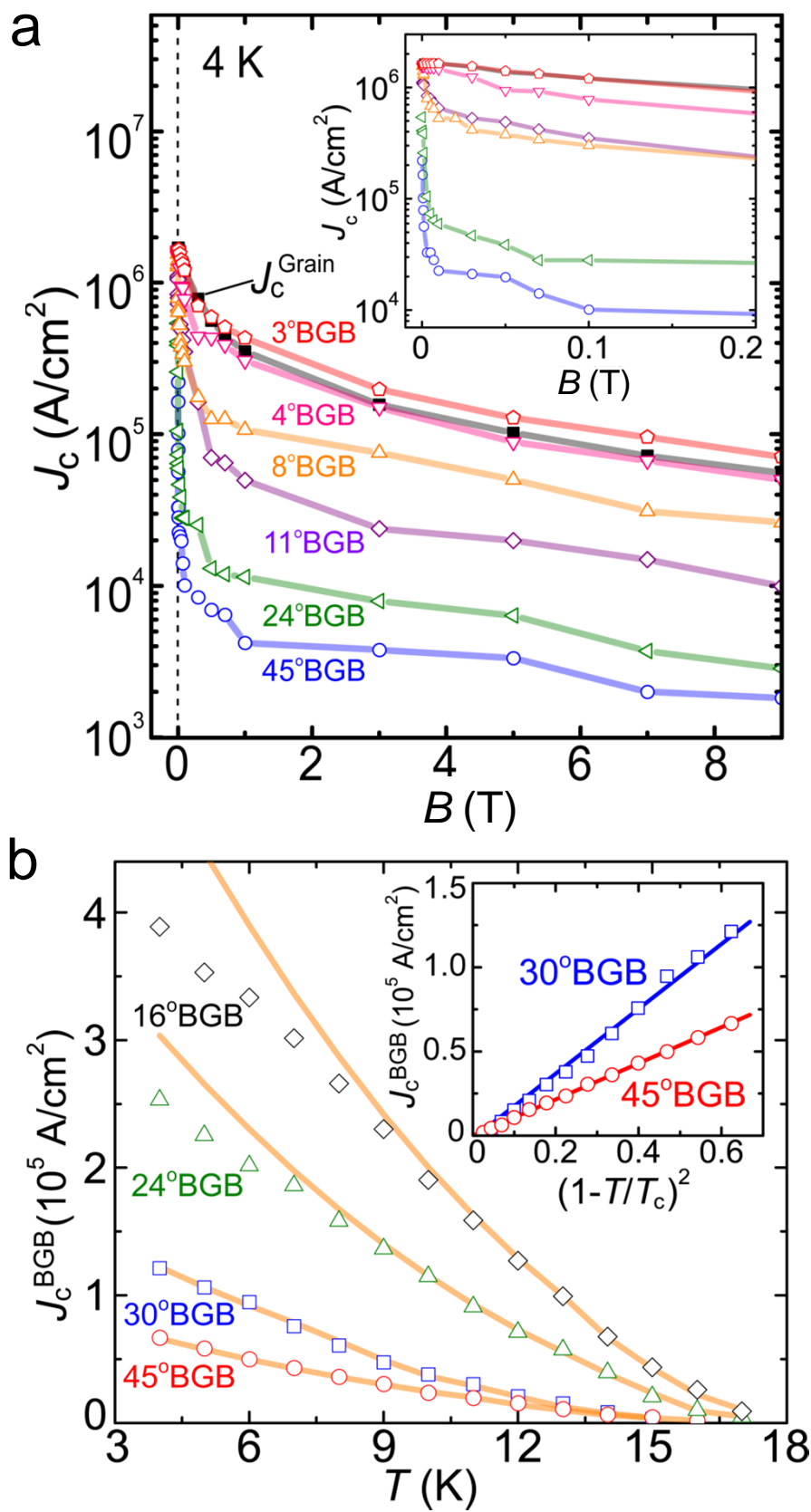

Figure 4